\documentclass[%
reprint,
 amsmath,amssymb,
 aps,
 prb,
floatfix,
showkeys
]{revtex4-2}

\usepackage{graphicx}
\usepackage{dcolumn}
\usepackage{bm}
\usepackage{hyperref}

\usepackage{xcolor}
\usepackage[normalem]{ulem}
\usepackage{comment}
\usepackage{lineno}
\usepackage{soul}
\begin{document}


\title{{Observation of Raman anomaly and characterization of magnetic phases} in van der Waals ferromagnet Fe$_5$GeTe$_2$}
\author{Sreelakshmi M. Nair, Aabhaas Vineet Mallik, R. S. Patel}
\email{rsp@goa.bits-pilani.ac.in}
 \affiliation{Department of Physics, Birla Institute of Technology \& Science 
Pilani - K K Birla Goa Campus, Goa - 403 726, India}

\date{\today}

\begin{abstract}
Two-dimensional (2D) van der Waals (vdW) ferromagnet Fe$_5$GeTe$_2$ has garnered significant interest due to its high Curie temperature (T$_C$), large saturation magnetization, and complex magnetic behavior arising, in part, from multiple inequivalent iron sites and vacancies. While several aspects of its complex magnetic and structural characteristics have been examined through careful experiments and first principles studies, much of it remains debatable. 
In this study, we present one of the first comprehensive temperature-dependent Raman spectrum for bulk Fe$_5$GeTe$_2$ and in the process reveal an interesting peak shift anomaly at 150 K.
We discuss the possible relationship of this Raman anomaly with the anomalous lattice expansion reported earlier for this material at around 110 K.
The impact of the anomalous lattice expansion on the magnetic anisotropy in this van der Waals material is also revealed by an isothermal magnetization analysis. 
These findings will prove crucial for the use of Fe$_5$GeTe$_2$ in high-performance spintronic devices.
\end{abstract}

\maketitle

\newpage

\section{Introduction}
Two-dimensional (2D) van der Waals (vdW) magnetic materials and their heterostructures have attracted significant research for exploring novel quantum phenomena and advancing spintronic devices. The weak interlayer coupling between the vdW layers facilitates easy mechanical exfoliation and effectively controls dimensionality in such systems. The strong magnetic anisotropy quenches the thermal fluctuations and stabilizes the long-range magnetic ordering in the 2D limit. In recent times, a wide range of vdW magnets with intrinsic magnetism in both bulk and thin layers have been discovered, including CrX$_3$ (X = I, Br, Cl)\cite{huang2017layer,mcguire2017magnetic,yang2023controlling}, MPS$_3$ (M = Mn, Fe, Ni) \cite{long2020persistence,lee2016ising,kim2019suppression}, CrXTe$_3$ (X = Si, Ge) \cite{liu2016critical,gong2017discovery} and Fe$_n$GeTe$_2$  ($n$ = 3, 4, 5) \cite{fei2018two,wang2023interfacial,fujita2022layer}. Among these materials
Fe$_3$GeTe$_2$ was the first metallic vdW magnet to exhibit a high Curie temperature (T$_C$) of 230 K in the bulk \cite{chen2013magnetic} and 130 K in the monolayer limit, along with strong perpendicular magnetic anisotropy \cite{fei2018two}.
Further, increasing the iron (Fe) concentration led to the discovery of new phases like Fe$_4$GeTe$_2$ and Fe$_5$GeTe$_2$, both with enhanced Curie temperatures. Fe$_4$GeTe$_2$ has a Curie temperature of 270 K \cite{seo2020nearly} while in Fe$_{5-x}$GeTe$_2$ (0 $<  x <$ 1), the Curie temperature can be varied between 270 and 330 K \cite{stahl2018van,may2019physical,alahmed2021magnetism}.

In addition to having a relatively high Curie temperature and large saturation magnetization (708 emu/cm$^3$) \cite{zhang2020itinerant}, the metallic ferromagnet Fe$_5$GeTe$_2$ also exhibits a very rich magnetic characteristics, making it the most promising vdW magnet for high-temperature spintronic device applications.
Cobalt substituted  Fe$_5$GeTe$_2$ has increased T$_C$, strong magnetic anisotropy, and even an antiferromagnetic state \cite{may2020tuning,tian2020tunable}. Recently, the magnetic ordering temperature (T$_C$) of epitaxially grown thin films of Fe$_5$GeTe$_2$  has substantially increased by self-intercalation \cite{silinskas2024self}. The reversible and non-volatile switching between two stable crystal structures in Fe$_5$GeTe$_2$ with distinct electronic structures has been observed \cite{wu2024reversible}. Furthermore, studies have demonstrated the operation of robust room-temperature lateral spin valves using heterostructures of Fe$_5$GeTe$_2$  and graphene, revealing a negative spin polarization at the interface \cite{zhao2023room}. The observation of topological magnetic spin structures such as skyrmions, merons, and anti-merons in  Fe$_5$GeTe$_2$ up to room temperature has introduced a new paradigm for fundamental research and spintronics applications \cite{klaui2022skyrmionic,casas2023coexistence,gao2020spontaneous}.

The presence of additional Fe sites and vacancies leads to intriguing magnetic behaviors in Fe$_5$GeTe$_2$, including multiple magnetic anomalies unlike other members of the Fe$_n$GeTe$_2$ family \cite{seo2020nearly,zhang2020itinerant}. In recent studies, these unconventional features of this material have been linked to charge density waves, spin reorientations, and spin glass formation \cite{wu2021direct,he2024spin,zhang2020itinerant}.
Despite significant theoretical and experimental progress, the exact origin of the magnetic anomalies in  Fe$_5$GeTe$_2$ remains debated. In particular, the cause of the anomalous lattice expansion at around 110 K \cite{may2019ferromagnetism} and its influence on the magnetic state of the system have not been settled comprehensively. Understanding the interplay of the lattice, spin and charge degrees of freedom of this metallic magnet is crucial for its practical spintronic applications. In this report, we investigated the intricate structural and magnetic properties of bulk Fe$_5$GeTe$_2$ crystal through X-ray diffraction (XRD), Raman spectroscopy, and various magnetization measurements. To the best of our knowledge this is the first temperature dependent Raman spectroscopy conducted on Fe$_5$GeTe$_2$ over the full range of temperature where this material exhibits non-trivial magnetic features \cite{Yu2025nonlinearhalleffect}. Interestingly, our Raman study reveals a hitherto unknown anomaly at 150 K but surprisingly does not show any pronounced signature of the anomalous lattice expansion around 110 K initially observed in neutron diffraction studies \cite{may2019ferromagnetism}. Moreover, while the magnetization (M) versus temperature (T) curve shows a distinct feature around 110 K suggesting a coupling between the structural and magnetic degrees of freedom, the Raman peak shift anomaly at 150 K leaves no distinct signature in the M(T) curve. We rationalize these observations in the light of thermal stability and charge transport measurements on this system reported in earlier studies \cite{may2019ferromagnetism,may2019physical,gao2020spontaneous} and implicate the vacancy ordering transition behind this phenomenology \cite{gao2020spontaneous, wu2021direct, gao2025electronicorderingdrivenflat, ly2021direct}. The interplay between the lattice and the spin degrees of freedom is further unveiled by our field-dependent magnetization measurements and the estimation of uniaxial magnetic anisotropy. 

\section{Results and Discussion}
\subsection{Structural Characterization of Fe$_5$GeTe$_2$ }
\begin{figure*}[h]
\centering
\includegraphics[width = 0.7\textwidth ]{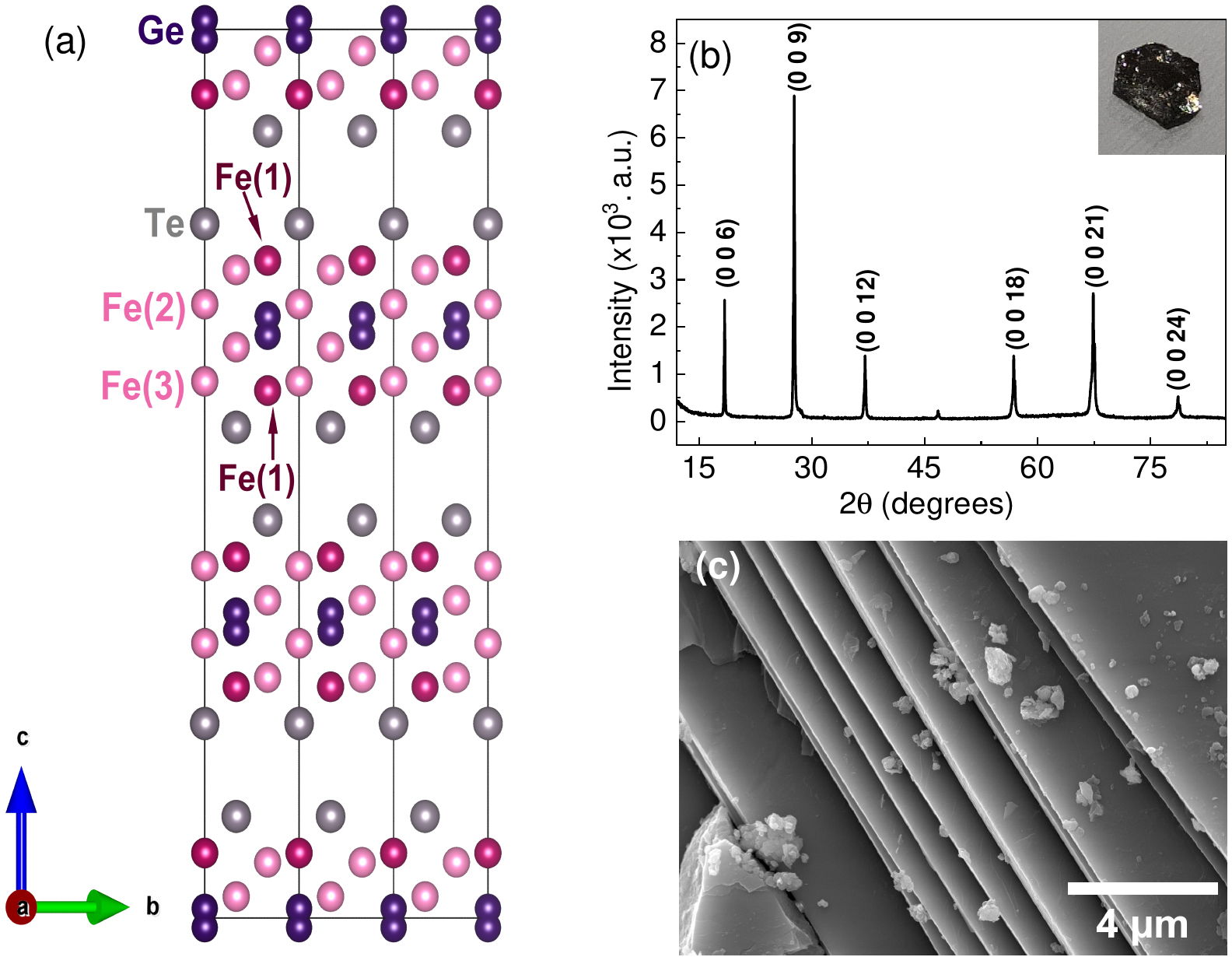}
\caption{\label{xrd}Structural charcterization of bulk Fe$_5$GeTe$_2$ single crystals.  (a) Schematic of the crystal structure of Fe$_5$GeTe$_2$ (b) Room temperature XRD of Fe$_5$GeTe$_2$ (c) FESEM image of bulk Fe$_5$GeTe$_2$ showing the vdW layers. }
\end{figure*}

Fe$_5$GeTe$_2$ belongs to the family of vdW metallic ferromagnets Fe$_n$GeTe$_2$ ($n$ = 3, 4, 5) where each layer is composed of Fe$_5$Ge slab sandwiched between Te layers. Unlike other Fe$_n$GeTe$_2$ systems, Fe$_5$GeTe$_2$ has a complex crystal structure due to split sites and vacancy disorder. Each slab comprises of nonequivalent Fe sites: Fe(1), Fe(2), and Fe(3), as indicated in Figure ~\ref{xrd}(a). The Fe(2) and Fe(3) sites are fully occupied, whereas the Fe(1) site in the outermost part of Fe$_5$Ge sublayer can occupy one of two possible split site positions. The occupancy of these Fe(1) split sites (either upward or downward) influences the positioning of the adjacent Ge atom, which shifts in the opposite direction (downward or upward, respectively) \cite{may2019physical,ly2021direct}.\par

Figure~\ref{xrd}(b) shows the x-ray diffraction (XRD) pattern of mechanically exfoliated Fe$_5$GeTe$_2$ crystals (commercially procured from HQ graphene) on Si/SiO$_2$ substrate. The XRD characterization of the sample was done using Bruker D8 Advance diffractometer with Cu K$_\alpha$ radiation $\lambda$ =1.54~\textup{~\AA} at room temperature. 
 The XRD pattern contains six (0 0 3$l$) peaks, specifically (0 0 6), (0 0 9), (0 0 12), (0 0 18), (0 0 21), (0 0 24). The presence of only (0 0 $l$) Bragg peaks demonstrates that the sample surface lies in the  $ab$ plane with the crystallographic $c -$axis oriented perpendicular to the surface \cite{zhang2023antisymmetric}. Figure~\ref{xrd}(c) shows the field emission scanning electron microscopy (FESEM) image of the bulk  Fe$_5$GeTe$_2$ crystal, illustrating the layered structure of vdW materials.

\subsection{Magnetic Characterization of Fe$_5$GeTe$_2$} \label{subsec:mag}
\begin{figure*}
\includegraphics[width = \textwidth ]{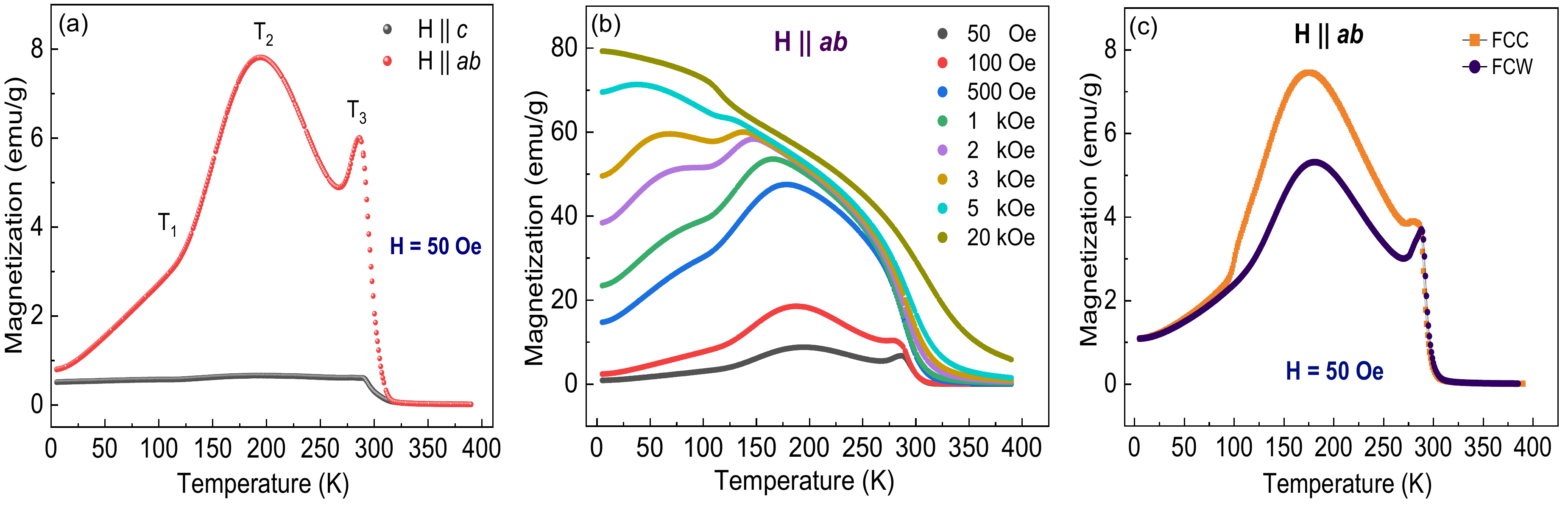}
\caption{\label{MT}Temperature-dependent magnetization measurements of bulk Fe$_5$GeTe$_2$ crystal. (a) Magnetization versus temperature M(T) for H $\parallel$ $ab$ and H $\parallel$ $c$ under an applied field H = 50 Oe. (b) M(T) curves for H $\parallel$ $ab$ at different magnetic fields. (c) Temperature-dependent Field cooled cooling (FCC) and Field cooled warming (FCW) magnetization for H $\parallel$ $ab$.}
\end{figure*}

We performed the magnetization measurements on the bulk Fe$_5$GeTe$_2$ crystal using the physical property measurement system (Quantum design PPMS Evercool II). The temperature-dependent magnetization M(T) of a zero-field-cooled (ZFC) sample was measured upon heating it from 5 to 390 K. Figure~\ref{MT}(a) shows the M(T) measurements under an applied magnetic field of H = 50 Oe 
for in-plane (H $\parallel$ $ab$ plane) and out-of-plane ( H $\parallel$ $c$ axis ) orientations. The out-of-plane magnetization is significantly 
smaller across all temperatures, whereas the in-plane (H $\parallel$ $ab$) data exhibits several distinctive features. The magnetization increases sharply upon cooling below 300 K marking the paramagnetic to ferromagnetic 
phase transition.  The Curie temperature (T$_C$) calculated from the minima of temperature derivative of the magnetization 
curve dM/dT is 294 K (Figure S2, Supporting Information). As shown in Fig.~\ref{MT}(a), three magnetic anomalies are observed in the ZFC curve marked by T$_1$, T$_2$, and T$_3$. The rise in magnetization below T$_C$ ceases at T$_3\sim 285$~K followed by a short but sharp drop. This anomaly at T$_3$ has been argued to arise from a competition between the Zeeman term, which dominates close to T$_C$, and the exchange interactions, which kick in once the average local moments exceed a threshold.  The exchange interactions in this system are highly anisotropic and predominantly ferromagnetic; however, owing to the local non-centrosymmetricity of the crystal, they also include the Dzyaloshinskii-Moriya (DM) term. The interplay of these two kinds of exchange interactions nucleates a helical magnetic state just below T$_3$ and leads to the observed drop in the average magnetization~\cite{ly2021direct}. As the strength of the external magnetic field increases the magnetization anomaly at T$_3$ diminishes and eventually disappears at a few hundred Oersteds [Fig.~\ref{MT}(b)]. This is not surprising and can be interpreted as the overwhelming of the DM interactions by the external magnetic field.

The next interesting feature in the temperature-dependent magnetization data (M(T)) for H $\parallel $ $ab$ in Fig.~\ref{MT}(a) is a hump at T$_2$ \cite{ly2021direct}. The increase in magnetization as the temperature decreases below T$_3$ (ignoring the dip just after T$_3$, which we have already discussed) can naturally be attributed to the presence of a varied set of inequivalent moments with predominantly ferromagnetic interactions \cite{ershadrad2022exchange}. Below T$_2$, however, the system exhibits a gradual decline in the average magnetization, which can be attributed to the formation of pronounced anti-ferromagnetic correlations arising from the weaker anti-ferromagnetic interactions between some of the moments \cite{ershadrad2022exchange}. This interpretation is corroborated by the expected dependence of T$_2$ on the external magnetic field. The magnetic hump occurs at T$_2$ $\sim$ 192~K for  H = 50~Oe and shifts to a lower temperature of  T$_2$ $\sim$ 167 K when the field is increased to H = 1 kOe [Fig.~\ref{MT}(b)]. Beyond H = 1 kOe, the hump at T$_2$ is strongly diminished, marking the suppression of the anti-ferromagnetic interactions by the external magnetic field. 
Interestingly, concomitant with the diminishing of the hump at T$_2$, the Curie temperature (T$_C$) also exhibits a non-monotonic behavior as a function of the external magnetic field. It decreases initially with increasing magnetic field, followed by a subsequent rise with further enhancement of the applied field. 
The T$_C$ is $\sim$ 294 K for H~=~50 Oe, and it gradually decreases to 292 K and 288 K for 100 Oe and 1 kOe, respectively. Above H = 1 kOe, the Curie temperature begins to rise and reaches a value of 310 K at H~=~20~kOe (Figure S3, Supporting Information). This is consistent with the suppression of the anti-ferromagnetic interactions by the external magnetic field $\sim$ 1 kOe.

Yet another interesting feature in the temperature-dependent magnetization data (M(T)) for H $\parallel $ $ab$ in Fig.~\ref{MT}(a) is the kink at T$_1$. The hump at T$_2$ and the kink at T$_1$ on the M(T) curve evolve differently with the applied magnetic field. Unlike T$_2$, the kink 
observed at T$_1$ $\sim$ 110 K for an applied field of  H~=~50~Oe persists even at higher magnetic fields with slight variation 
in the transition temperature [Fig.~\ref{MT}(b)]. Many theories have been proposed to explain these unconventional magnetic behaviors at low temperatures, 
including transitions to spin glass state, formation of charge density waves (CDW), and Heavy Fermion (HF) state, etc. \cite{zhou2024identifying,gao2020spontaneous,zhang2020itinerant}.
However, neutron diffraction studies by A. F. May et al.~\cite{may2019ferromagnetism} strongly suggest that the anomaly at T\(_1 \) is associated with the change in lattice parameters,  which sets in at about 150 K and culminates in a sharp anomalous expansion along the $a$-axis at $\sim$110~K.

Field-cooled cooling (FCC) and field-cooled warming (FCW) magnetization measurements were performed on the sample as a function of temperature under an applied field H~=~50~Oe, parallel to the $ab$ plane [Fig.~\ref{MT}(c)]. All three anomalies observed in the ZFC data [Fig.~\ref{MT}(a)] are also present in the FCC and FCW curves. Remarkably, a thermal hysteresis loop is found to develop around these anomalies, opening at 288 K $\gtrsim$ T$_3$ and closing at 95~K~$\lesssim $ T$_1$, with a temperature span of  $\Delta$T $\approx$ 193 K where the moments in the system fail to thermalize \cite{may2019physical}. While the absence of hysteresis above 288 K does not rule out the absence of thermalization (as the magnitude of magnetization itself is strongly suppressed due to thermal fluctuations), its absence below 95 K is a telltale sign of the emergence of strong coupling between the phonons and the magnetic moments in the system in this temperature range. This is consistent with the lattice contraction along the $c$-axis that the system undergoes around the same temperature \cite{may2019ferromagnetism}.

\begin{figure*}[h]
\centering
\includegraphics[width = 0.8\textwidth ]{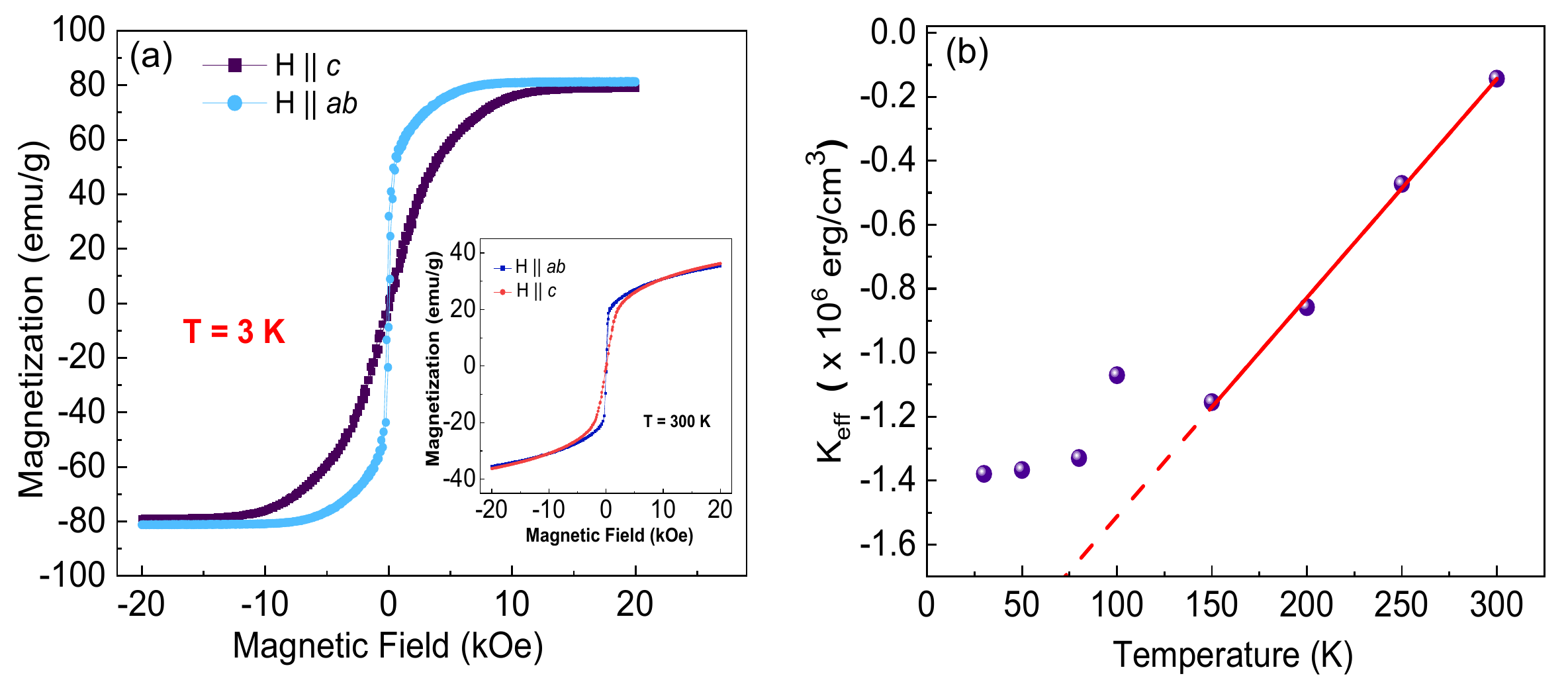}
\caption{\label{MH}(a) Isothermal magnetization curves M(H) for H $\parallel$ $ab$ and H $\parallel$ $c$ at 3 K. Inset shows the M(H) curve at 300 K. (b) Temperature dependence of the uniaxial magnetic anisotropy \( K_{\text{eff}} \). The red line is a straight line joining the data points above 125 K.}
\end{figure*}

Figure~\ref{MH}(a) represents the isothermal magnetization measurements M(H) as a function of the applied magnetic field along the $c$ - axis and $ab$ plane at 3 K. The saturation field for H $\parallel$ $ab$ is smaller than the H $\parallel$ $c$, confirming that the easy axis lies in the $ab$ plane. The inset of Fig.~\ref{MH}(a) shows the M(H) data at 300~K for the two orientations of the magnetic field, suggesting that a small anisotropy exists even at higher temperatures. The low coercive field value of H$_{ab}$ = 50 Oe for H~$\parallel$~$ab$ and H$_c$ = 109 Oe for H $\parallel$ $c$ indicates the soft ferromagnetism in Fe$_5$GeTe$_2$, similar to Fe$_3$GeTe$_2$ \cite{liu2017critical}.
To further investigate the properties of the Fe$_5$GeTe$_2$ crystal, the change in uniaxial magnetocrystalline anisotropy is estimated at various temperatures. The uniaxial anisotropy K$_\mathrm{eff}$ is calculated by taking the area difference between the out-of-plane and in-plane magnetization curves \cite{dhakal2010magnetic,singh2022study}.
The negative sign indicates strong in-plane anisotropy in the crystal. Fig.~\ref{MH}(b) demonstrates the variation of magnetic anisotropy in the temperature range 25$-$300 K. The deviation from the high temperature linear behavior (red line in Fig.~\ref{MH}(b)) at temperatures below $\sim$~150 K is consistent with contraction along the $c$-axis and the emergence of strong magnetoelastic coupling below T$_1$, as discussed in the previous paragraph. In contrast to earlier studies of Fe$_5$GeTe$_2$, which report a rotation of magnetic anisotropy from the easy axis to the easy plane with a decrease in temperature  \cite{may2019physical, seo2020nearly, lv2022controllable}, in our case, the anisotropy lies in the $ab$-plane across all temperatures, as shown in Fig.~\ref{MH}(b).

\begin{figure*}[h]
\centering
\includegraphics[width = 0.7\textwidth ]{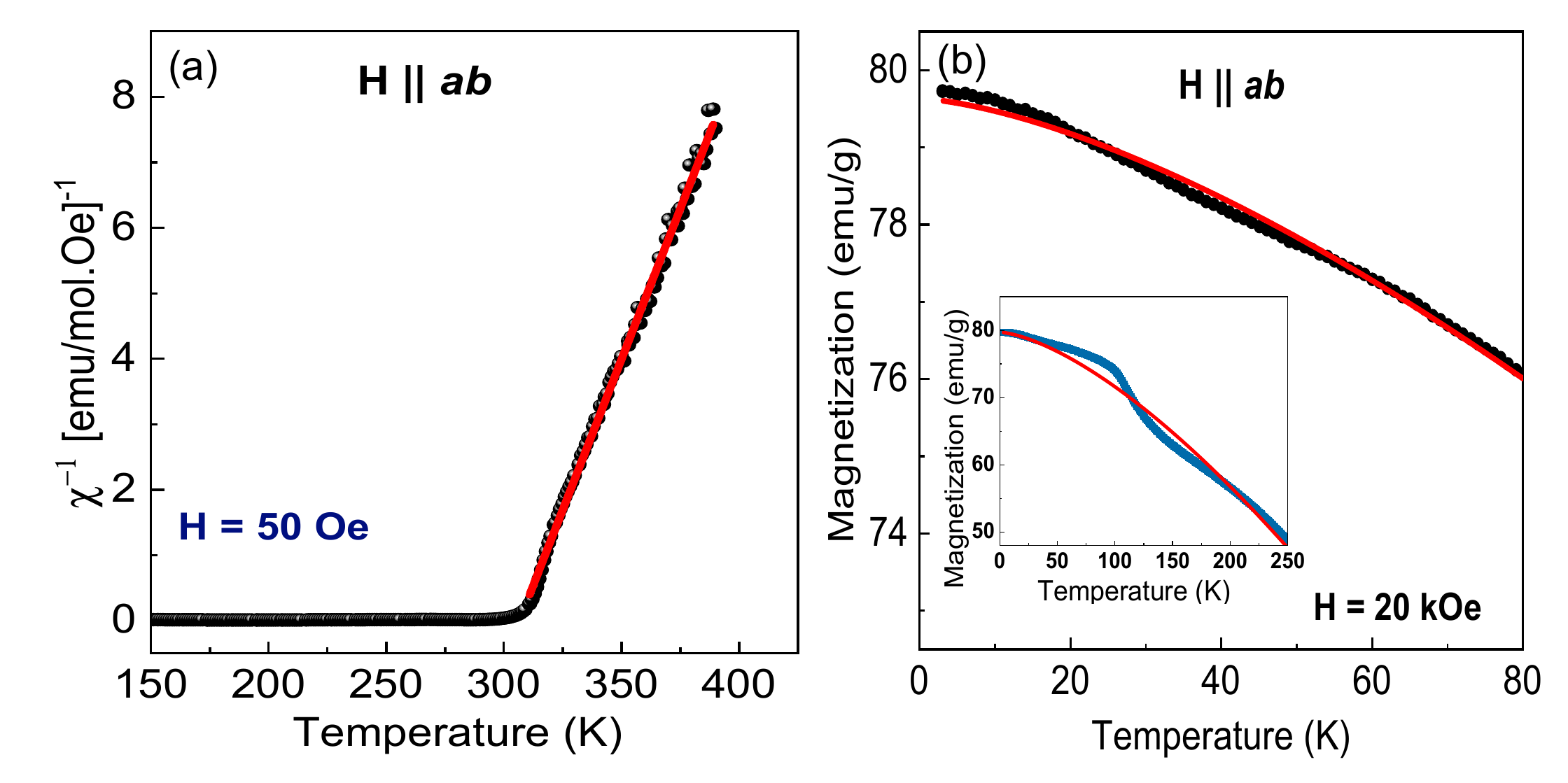}
\caption{\label{fitting}(a) Reciprocal molar susceptibility vs temperature for  H = 50 Oe applied parallel to the $ab$ plane. The red line indicates the Curie-Weiss fitting in the temperature range 310$-$390 K. (b)~M(T) curve above saturation magnetic field H = 20 kOe fitted by standard Bloch spin-wave model in the low temperature region (5$-$80 K). The inset shows the Bloch fitting in the wider temperature region (5$-$250 K) [solid red line represents the fitting].}
\end{figure*}

Figure~\ref{fitting}(a) shows the reciprocal susceptibility ($1/\chi$) versus temperature plot for H~=~50~Oe applied parallel to the $ab$ plane (see Figure S4, Supporting Information). The characteristic Curie-Weiss temperature ($\theta$) and the effective magnetic moment ($\mu_{\mathrm{eff}}$) are calculated by fitting the Curie-Weiss law [$1/\chi = $ ($T-\theta$)$/C$], where \textit{C} is the Curie constant, in the high-temperature region (310$-$390 K) of the ZFC data. The temperature range selected is well beyond the critical region associated with the magnetic phase transition at $T_C$. From the fitting, the Curie-Weiss temperature ($\theta$) is determined to be $\theta$$_a$$_b$ = 306 K for H $\parallel$ $ab$ and $\theta$$_c$~=~303 K for H $\parallel$ $c$ cases, both of which are higher than the transition temperature T$_C$. The positive $\theta$ values indicate the dominant ferromagnetic interactions between the moments in the system. The effective magnetic moment ($\mu_{\mathrm{eff}}$) is calculated to be  1.87 $\mu_B$/Fe for H~$\parallel$~$ab$  and  1.97 $\mu_B$/Fe for H $\parallel$ $c$  orientations. The observed similarities between the ($\mu_{\mathrm{eff}}$) values strongly suggest the nearly isotropic paramagnetic behaviors at high temperatures in the sample akin to Fe$_3$GeTe$_2$  \cite{liu2017critical}. 

To further understand the ferromagnetic behavior of bulk Fe$_5$GeTe$_2$, the M(T) data at H~=~20 kOe is fitted using the standard 
Bloch spin wave model \cite{Bloch1930, Kittel_SSP} 
$M = M_0$($1 - B T^{3/2}$)] where \( M_0 \) is the saturation magnetization at 0 K and  \( B \) is the Bloch constant. Fig.~\ref{fitting}(b) shows the Bloch fitting for  H $\parallel$  $ab$ case at low temperatures, 5 $-$ 80 K, much below the Curie temperature (see Figure S5 of Supporting Information for H $\parallel$ $c$). 

 The inset of Fig.~\ref{fitting}(b), represents the the same Bloch fitting to a higher temperature, up to 250 K. Overall the magnetization data still follows the Bloch law but also a unique deviation is observed  about 110 K.  This can be attributed to the sharp change in the lattice parameters previously reported in neutron diffraction observations \cite{may2019ferromagnetism}. Additionally, the spin wave stiffness constant (\( D_{\text{spin}} \)) calculated from the Bloch model is \( 3.95 \times 10^{-22} \, \text{J nm}^2 \), and the value of the exchange stiffness (\( A_{\text{stiff}} \)) constant at 0 K is \( 2.13 \, \text{pJ/m} \) (Section S6, Supporting Information).

\subsection{Raman spectroscopy study of Fe$_5$GeTe$_2$}
\begin{figure*}
\centering
\includegraphics[width = 1\textwidth ]{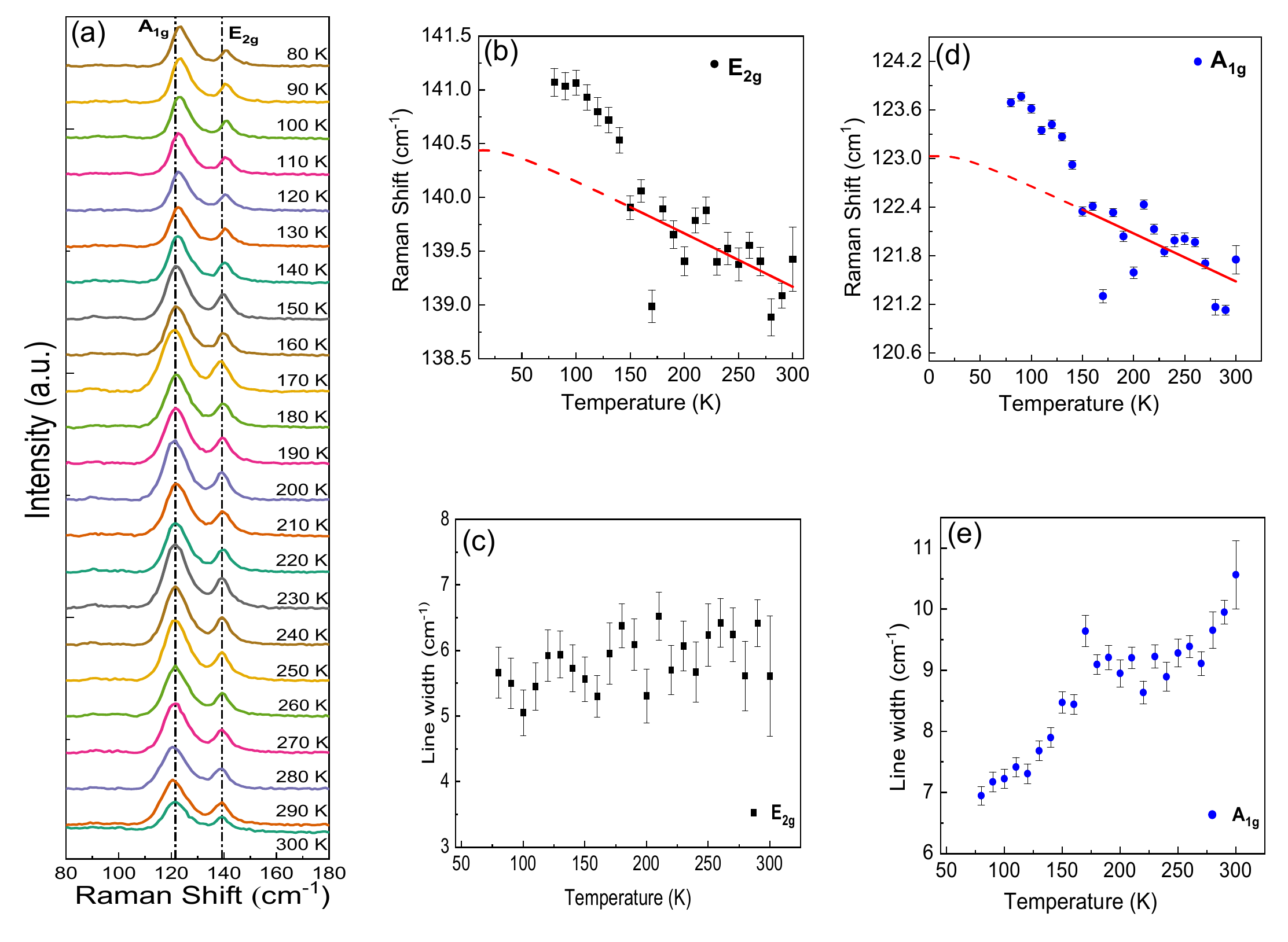}
\caption{\label{raman}Raman spectroscopy of bulk Fe$_5$GeTe$_2$ crystal. (a) Temperature dependent Raman spectra from 80 to 300 K. The black dash-dotted lines shows the position of the peaks at 300 K. It is clearly visible that the peak position is shifted towards higher energy at low temperatures. (b) and (d) Temperature evolution of Raman modes  E$_{2g}$ and $A_{1g}$, respectively. Red solid lines represent the fitting using a model for anharmonic phonons as mentioned in the text and dashed line is extrapolation to lower temperatures. (c) and (e) Temperature dependence of peak-widths (FWHM) of  E$_{2g}$ \& $A_{1g}$ peaks respectively.}
\end{figure*}

The bulk Fe$_5$GeTe$_2$ crystals were characterized by Raman spectroscopy using  LabRAM HR Horiba Raman spectrometer with an excitation laser of wavelength 532 nm. The Raman spectra of the crystal show two characteristic peaks at all temperatures. The Figure S1 of supporting information represents the Raman spectrum at room temperature of bulk Fe$_5$GeTe$_2$. It shows two peaks at 121.5 $ \text{cm}^{-1} $  and 139.2 $ \text{cm}^{-1} $ which corresponds to $A_{1g}$ and $E_{2g}$ modes respectively similar to Fe$_3$GeTe$_2$  \cite{liang2023facile,weerahennedige2024effects, Yu2025nonlinearhalleffect}. A temperature-dependent Raman spectroscopy measurement was performed on the sample from 300 to 80 K with a step size of 10 K to investigate the changes in the behavior of the phonon modes associated with the changes in the magnetic and electronic character of the system. The $A_{1g}$ peak consistently exhibits higher intensity than the $E_{2g}$ peak at all temperatures. Initially, as the temperature decreases from 300 K, the peak positions gradually shift towards higher frequencies, but below  150 K, both peaks shift abruptly by $\sim$ 2 cm\(^{-1}\)
[Fig.~\ref{raman} (a)]. With decreasing temperature the \(A_{1g}\) mode shifts from 121.5 cm\(^{-1}\) at 300 K to 123.6~cm\(^{-1}\) at 80 K. In the same way the \(E_{2g}\) mode shifts from 139.2 cm\(^{-1}\) for T = 300 K to 141.3~cm\(^{-1}\) at T~=~80~K. For a detailed analysis each spectrum was fitted using a sum of Lorentzian line shapes. The temperature evolution of the extracted peak positions were plotted as a function of temperature in Fig.~\ref{raman}(b) and Fig.~\ref{raman}(d). As shown in these figures the phonon frequencies of both the peaks increase with decrease in the temperature. The experimental data in the temperature range 150 $-$ 300 K was fitted with the equation, $$\omega_{\text{anh}} = \omega_{0} - C \left( 1 + \frac{2}{\exp\left(\hbar \omega_0/2 k_B T\right) - 1} \right),$$  where $\omega_{0}$ and C are adjustable parameters, to account for anharmonic effects \cite{balkanski1983anharmonic,bhadram2013spin}. The red solid line in figure represents the best fit in the temperature range 150 $-$ 300 K. Below 150~K, the extrapolated red dashed line shows significant deviation from the experimental data.

\begin{figure*}
\centering
\includegraphics[width = 0.9\textwidth ]{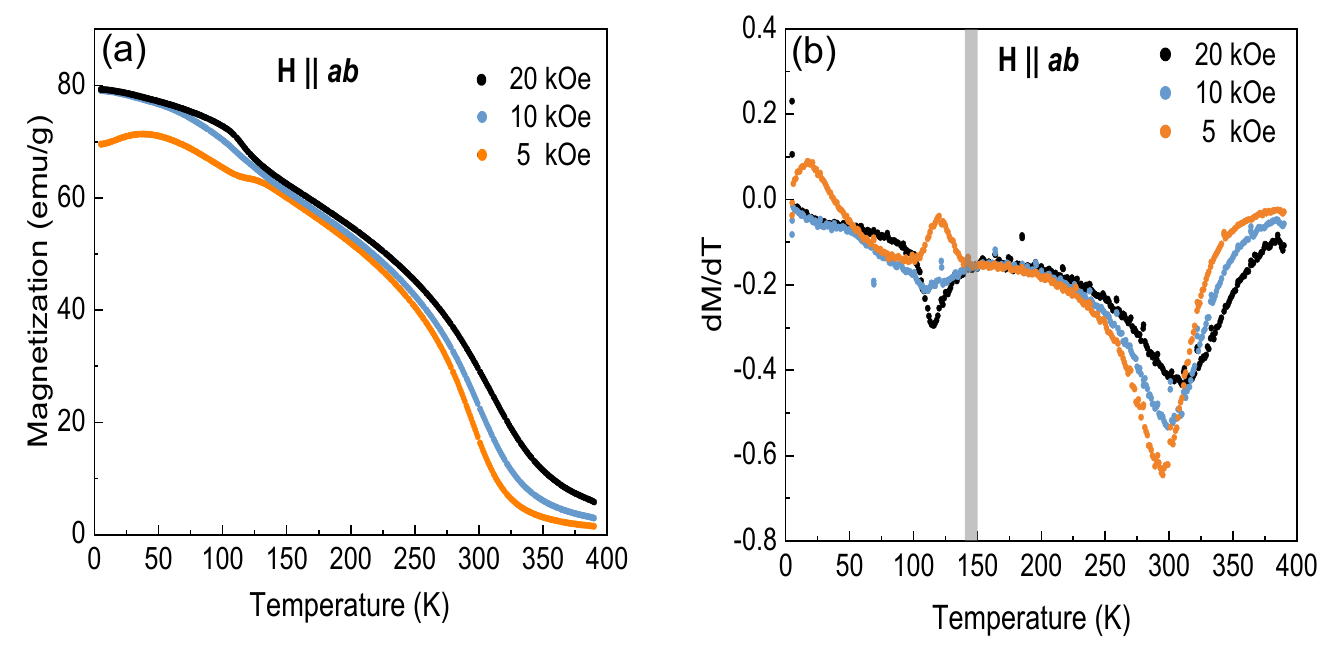}
\caption{\label{dM_dT} (a) Magnetization versus temperature plots for H = 5, 10 and 20 kOe applied parallel to the $ab$ plane. (b) Corresponding dM/dT curves as function of temperature. Grey region represents the the feature at T = 150 K described in the text.}
\end{figure*}

It is interesting to note that the peak shift anomaly around 150 K in the Raman spectra of the system has no pronounced effect on the magnetic character of the system detailed in the section~\ref{subsec:mag}, except for possibly in the uniaxial anisotropy K$_\mathrm{eff}$ (see Fig.~\ref{MH}(b)) and another subtle signature which we now highlight. In Fig.~\ref{dM_dT} (b) we plot dM/dT as a function of temperature for three different   magnetic fields (5, 10 and 20 kOe) applied along  $ab$ plane. These fields are large enough to essentially overwhelm the microscopic exchange interactions between the magnetic moments of the system and force a spin polarized state (see Fig.~\ref{MH}(a)). The corresponding M(T) curves have been reproduced in Fig.~\ref{dM_dT}(a) for better visualization. Remarkably, while the three dM/dT curves in Fig. 6(b) are generally different
from each other, they all come together to form a distinct feature around 150 K. In the light of the observations made in Fig.~\ref{MT}(c) and in Ref.~\onlinecite{may2019physical} about the lack of significant magnetoelastic coupling for temperatures between T$_1$ and T$_3$, it is not surprising that what is pretty pronounced in the Raman data leaves only soft signatures in the magnetization data.

We now discuss the signatures of the Raman anomaly around 150 K in the published lattice parameter data obtained by neutron diffraction and the available charge transport data for the system. Let us first consider the lattice parameter data presented in Fig.~2(d) of Ref.~\onlinecite{may2019ferromagnetism}. While the discontinuity in the slope of the $c/a$ vs T curve occurs close to T$_1 \sim$~110~K and is associated with the anomalous lattice expansion, the slope of this curve starts to change conspicuously at around 150 K from its nearly constant value at high temperatures. Next let us consider the charge transport data presented in Fig.~4(a) of Ref.~\onlinecite{zhang2020itinerant}. Here again the dR/dT vs T curve gets a sharp upturn close to 150 K, when approached from the high temperature side, and attains a pronounced peak only at T$_1$. Similarly, the normal Hall co-efficient data reported in the inset of Fig. 3(c) of Ref.~\onlinecite{gao2020spontaneous} also shows a significant upturn close to 150 K, when approached from the high temperature side, and attains a maximum at T$_1$.

It is apparent from the above discussion that the observed  anomaly in Fe$_5$GeTe$_2$ around 150 K from Raman measurements   exhibits only soft signatures in the magnetic, electronic and static lattice parameters of the system. This phenomenology suggests that what transpires at 150 K is not driven by the magnetic, electronic or static lattice parameter changes in the system. Rather, it is likely to be an independent and subtle transition which, by its coupling with the magnetic moments in the system, eventually nucleates the first order transition at T$_1$ \cite{may2019ferromagnetism}. Interestingly, this system does have a degree of freedom which can host such a transition at 150 K, namely the split position of Fe(1) sites in each unit cell. It is known that in the annealed samples of Fe$_5$GeTe$_2$ the Fe(1) sites prefer to be in the disordered phase close to T$_C$ and at some temperature between T$_3$ and T$_1$ they make a transition to a short range order phase in the $ab$-plane \cite{may2019ferromagnetism,may2019physical,ly2021direct,wu2024reversible}. However, to the best of our knowledge the actual temperature where this transition takes place has thus far not been established precisely.  We propose that this vacancy ordering transition in Fe$_5$GeTe$_2$ occurs around 150 K and can be verified directly via Scanning Tunneling Microscopy (STM) observations carried out over a range of temperatures between T$_3$ and T$_1$.

A crucial hint regarding the nature of the ordering transition at 150 K is obtained by looking at the variation of the width of the Raman peaks across this temperature. The widths of the $E_{2g}$  and $A_{1g}$ peaks estimated from our Lorentzian fits (FWHM) at different temperatures are plotted in Figure~\ref{raman}(c) and (e), respectively. Interestingly, while the width of the $A_{1g}$ peak reduces markedly below 150 K and is consistent with an ordering transition, the width of the $E_{2g}$ peak remains essentially unchanged throughout the plotted temperature range. This difference between the width of $A_{1g}$ and the $E_{2g}$ peaks can be attributed to the difference in their symmetries (for a detailed discussion, see, for example, Ref.~\onlinecite{Yu2025nonlinearhalleffect}) and the symmetry of the order parameter which sets in at 150 K. The Raman active $A_{1g}$ mode of Fe$_5$GeTe$_2$ is associated with vibrations along the $c$-axis and thus, a reduction in its peak width indicates ordering of the Fe(1) sites along the $c$-axis. Furthermore, the absence of width reduction of the $E_{2g}$ peaks, which is associated with vibrations in the $ab$-plane, indicates that the Fe(1) sites in the $ab$-plane prefer to remain largely disordered. This is consistent with the short range positional order in the Fe(1) sites along the $ab$-plane, reported earlier \cite{ly2021direct}. The scenario of ordering of the Fe(1) sites along the c-axis is also consistent with the fact that it leaves only a soft signature in the electronic and magnetic properties of the system which are largely dominated by interactions in $ab$-plane in this van der Waals material.

\section{Conclusion}
In conclusion, we have systematically analyzed the magnetic characteristics of Fe$_5$GeTe$_2$ and its correlation with the structural features of the crystal. We have presented the first comprehensive Raman spectra for this material in the temperature range of 80 $-$ 300 K. This allowed us to uncover a peak shift anomaly at 150 K which was insofar unknown. We propose this anomaly in the Raman peak shift is a result of the ordering in the Fe(1) sites among the unit cells of the system. We also argue that, this anomaly, because of its coupling with the magnetic moments in the system would eventually lead to the lattice anomaly at 110 K.
Isothermal magnetization measurements and magnetic anisotropic calculations confirm that the sample favors in-plane magnetization. Furthermore, the parameters including characteristic  Curie-Weiss temperature ($\theta$),  effective magnetic moment ($\mu_{\mathrm{eff}}$), spin wave stiffness constant (\( D_{\text{spin}} \)) and exchange stiffness (\( A_{\text{stiff}} \)) are determined to understand the magnetic nature of the system better.
Our findings while confirming the intricate interplay of different exchange interactions in this van der Waals magnet, uncover the interesting possibility of tuning the magnetoelastic coupling with temperature making it a very potent candidate component of future spintronic devices.

\textbf{Supporting Information} \\
Supporting Information is available online.

\begin{acknowledgments}
 One of the authors, Sreelakshmi M Nair, thanks BITS Pilani for the financial support. We
acknowledge the Department of Physics, BITS Pilani 
K K Birla Goa Campus and DST, Govt
of India for DST FIST grant number SR/FST/PS-I/2017/21. We would like to thank the Central Sophisticated Instrumental Facility (CSIF), BITS Goa, for their assistance with XRD and Raman measurements.   
\end{acknowledgments}

\bibliography{FGT}

\end{document}